\documentclass[aps,prl,preprint,showkeys,groupedaddress]{revtex4-1}

\usepackage{amstext}
\usepackage{amsmath}
\usepackage{graphicx}
\usepackage{bm}
\usepackage{color}

\usepackage[normalem]{ulem}

\definecolor{bordeaux}{rgb}{0.65, 0.0, 0.05}
\definecolor{BORDEAUX}{rgb}{0.65, 0.0, 0.05}

\begin{document}
\title{Self-organization principles of intracellular pattern formation}
\author{J. Halatek. F. Brauns, and E. Frey}
\email{frey@lmu.de}
\affiliation{Arnold--Sommerfeld--Center for Theoretical Physics and Center for NanoScience, Department of Physics, Ludwig-Maximilians-Universit\"at M\"unchen, Theresienstra{\ss}e 37, D-80333 M\"unchen, Germany}

\date{\today}

\begin{abstract}

Dynamic patterning of specific proteins is essential for the spatiotemporal regulation of many important intracellular processes in procaryotes, eucaryotes, and multicellular organisms. The emergence of patterns generated by interactions of diffusing proteins is a paradigmatic example for self-organization. 
In this article we review quantitative models for intracellular Min protein patterns in \textit{E. coli}, Cdc42 polarization in \textit{S. cerevisiae}, and the bipolar PAR protein patterns found in \textit{C. elegans}. 
By analyzing  the molecular processes driving these systems we derive a theoretical perspective on general principles underlying self-organized pattern formation.
We argue that intracellular pattern formation is not captured by concepts such as ``activators'', ``inhibitors'', or ``substrate-depletion''. Instead, intracellular pattern formation is based on the redistribution of proteins by cytosolic diffusion, and the cycling of proteins between distinct conformational states. Therefore, mass-conserving reaction-diffusion equations provide the most appropriate framework to study intracellular pattern formation. 
We conclude that directed transport, e.g. cytosolic diffusion along an actively  maintained cytosolic gradient, is the key process underlying pattern formation. Thus the basic principle of self-organization is the establishment and maintenance of directed transport by intracellular protein dynamics.   
%
%
%
%
%
%

\end{abstract}



\maketitle

\section{Introduction}
In biological systems self-organization refers to the emergence of spatial and temporal structure. Examples include the structure of the genetic code, the structure of proteins, the structure of membrane and cytoplasm, or those of tissue, and connected neural networks. 
On each of these levels, interactions resulting from the dynamics and structural complementarities of the system's constituents bring about the emergence of biological function. Biological systems are the perfect example for the Aristotelian notion that ``the whole is more than the sum of it's parts''. For centuries this phrase expressed nothing more than a vague intuition that some set of organizational principles must underlie the complex phenomena we observe around us. Owing to the advances in quantitative biology and theoretical biological physics in recent decades, we have begun to understand how biological structure and function originates from fundamental physical principles of self-organization. While we are not yet in a position to define any \emph{universal} physical principles of self-organization in general, we are now able to identify recurring themes and principles in particular, important areas like intracellular pattern formation. This will be the main focus of this review article.

\begin{figure}
	\includegraphics{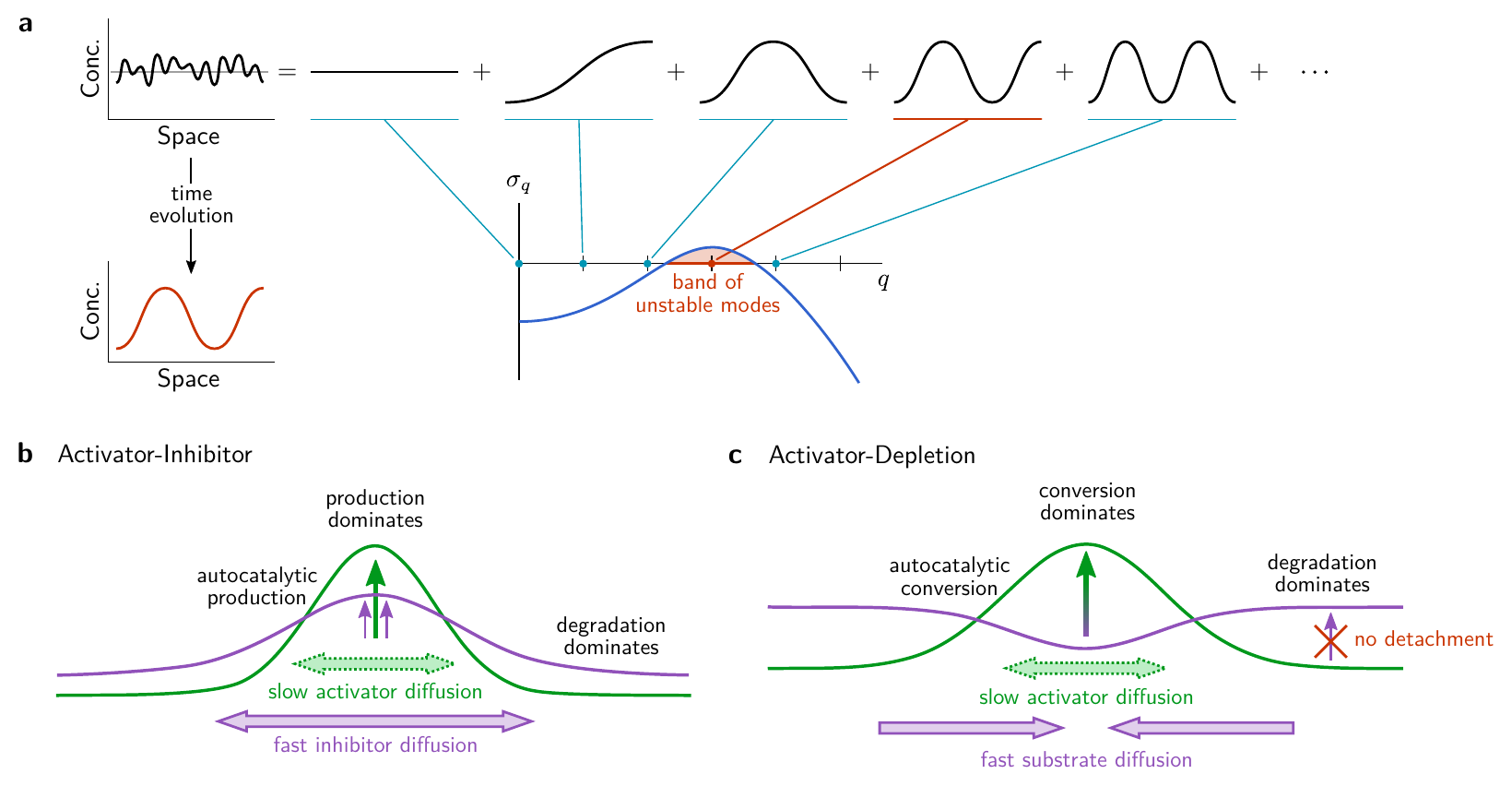} 
	\caption{Turing's general linear stability analysis and heuristic ad-hoc interpretations in terms of the activator--inhibitor picture based on the production and degradation of reactants.
	(a) Any random perturbation (black line) of a spatially uniform state (gray line) can be decomposed into Fourier modes. Linear stability analysis yields the growth rates $\sigma_q$ of the amplitude of all modes $q$. This is represented in the dispersion relation (blue line $\sigma_q$). Unstable modes (marked red) grow in amplitude an determine the pattern emerging out of the random perturbation during the incipient time evolution.
	(b) The activator--inhibitor model is based on autocatalytic production of a slowly diffusing activator, which in turn stimulates the production of a fast-diffusing inhibitor that suppresses autocatalytic activator production. Both activator and inhibitor are subject to degradation. The faster diffusion of the inhibitor leads to the formation of an inhibition zone in which degradation dominates over activator (and inhibitor) production. 
	(c) In the activator--depletion model the inhibitor is replaced by a substrate that is subject to degradation, and autocatalytic activator production is replaced by the autocatalytic conversion of substrate into activator. The rate of conversion is limited by the available substrate. Heuristically, this conversion could be equated with the attachment of cytosolic proteins to the membrane. However, the reverse process (detachment) is not taken into account. Both substrate and activator are steadily degraded and are produced at a finite rate. If the activator density is too low, the conversion process is suppressed and the degradation process dominates, as in the activator--inhibitor model. }
	\label{fig:activator-inhibitor}
\end{figure}

The generic equilibrium state of any diffusion process is spatially uniform as diffusion removes spatial gradients in chemical concentration. Self-organized pattern formation implies that this equilibrium can be destabilized, such that an initially uniform system evolves towards a non-uniform steady state --- a pattern \cite{Turing1952, Cross1993}.
Historically, the field of self-organized  pattern formation in chemical systems was initiated by Alan Turing in 1952 \cite{Turing1952}. 
In his seminal article on ``\textit{The chemical basis of morphogenesis}'', Turing showed that the interplay between molecular diffusion and chemical interactions can give rise to an instability of the spatially uniform state. 
His general finding was that in a system with multiple reacting components diffusing laterally on different time scales, the diffusive coupling itself can cause an instability even if the system is in a stable chemical equilibrium.
Turing was the first to introduce a linear stability analysis for reaction-diffusion systems. To give the reader an impression of the generality of his ideas let us briefly summarise the underlying mathematical concepts:  The initial idea is that any random perturbation of a uniform steady state can be decomposed in Fourier modes $\sim \cos(qx)$ (Figure \ref{fig:activator-inhibitor}a). As long as amplitudes are small, each of these modes grows or decays exponentially $\sim \exp(\sigma_q t)\cos(qx)$, depending on the sign of the growth rate $\sigma_q$ (or more precisely the real parts $\text{Re}[\sigma_q]$). By linear stability analysis one computes the growth rates for modes with any wavenumber $q$. Turing found that the interaction between chemical reactions and molecular diffusion can give rise to bands of unstable modes with positive growth rates, i.e. situations where some modes with particular wavelengths are amplified out of a random perturbation (Figure \ref{fig:activator-inhibitor}a). This gives rise to pattern formation. In the following we will refer to the pattern forming instability as \emph{lateral instability} since it originates from lateral diffusive coupling.

As proof of principle Turing demonstrated this stability analysis for a general reaction-diffusion system with two chemical components, but also discussed (oscillatory) cases with three components.
For 20 years his results received very little attention. It was only in 1972 when Segel and Jackson \cite{Segel1972} first interpreted Turing's linear stability analysis of the two-component model, while Gierer and Meinhardt \cite{Gierer1972a} in the same year proposed a number of specific two-component models and coined the terms ``activator'', ``inhibitor'', and ``activator--inhibitor mechanism'' in this context. 

Unfortunately, nowadays the terms ``activator--inhibitor mechanism'' and ``Turing instability'' are often thought to refer to identical concepts, despite the fact that the ``activator--inhibitor mechanism'' only represents a particular interpretation of Turing's proof of principle analysis that is specific to some but not all two-component models. Furthermore, note that Turing's general idea of a lateral instability is in fact not even limited to a particular number of chemical components.
In the literature the ``activator--inhibitor mechanism'' is usually considered as a combination of ``short range activation'' and ``long range inhibition'' or of ``local activation'' and ``lateral inhibition'' in order to convey the following heuristic picture \cite{Meinhardt2000,Meinhardt2008,Cross1993}:

Consider two chemical components. First -- the (short range) activator -- enhances its own \emph{production} in some autocatalytic fashion such that its concentration can increase exponentially. If the diffusion coefficient of this component is small -- any concentration peak will only slowly disperse in the lateral direction. Secondly, the (long range) inhibitor, which is also produced by the activator, but has a much large diffusion coefficient. Hence, it does not accumulate locally with the activator but disperses laterally, where it inhibits the action of the activator (see Fig.~\ref{fig:activator-inhibitor}b).
It is crucial to realize that this mechanisms is merely a heuristic interpretation of a formal linear stability analysis presented by Turing. A quite common misunderstanding in the biological literature is that pattern formation \emph{requires} an activator and an inhibitor. This does not in any way follow from the analysis by Turing \cite{Turing1952}, Segel and Jackson \cite{Segel1972}, Gierer and Meinhardt \cite{Gierer1972a,Meinhardt2000,Meinhardt2008}, or any other analysis -- activator--inhibitor models are simply mathematically idealized examples of pattern forming systems. Moreover, the underlying interpretation is actually  restricted to systems with only \emph{two} interacting chemical components. Note that ``chemical component'' does not refer to a protein species, but to the \emph{conformational state} of a protein that determines its interactions with specific (conformations of) other proteins. Clearly, protein interaction networks include many conformational states -- not just two \cite{Frey:2018a}.

Furthermore, the activator--inhibitor interpretation inextricably links chemical properties (e.g. autocatalytic action) to the diffusibility of the components (e.g. short range activation). 
However, in the context of intracellular protein pattern formation the general distinction between diffusibilities is that between membrane-bound and cytosolic (conformational) states. 
Accordingly, membrane-bound protein conformations would have to be considered as activators in the activator--inhibitor picture, and cytosolic protein conformation as inhibitors, respectively. There are many reasons why this picture is not applicable to intracellular protein dynamics -- the most glaring discrepancy is that proteins are not produced autocatalytically on the membrane, which is the major (implicit) assumption underlying all activator--inhibitor interpretations.
As we will discuss in detail below, intracellular protein pattern formation is generically independent of protein production and degradation (cf. \cite{Shamir:2016a}), and intracellular protein dynamics are generically driven by the cycling of proteins between membrane-bound and cytosolic conformations.

Another interpretation of Turing's mathematical analysis of two-component systems, which appears to take these considerations into account, is the ``activator--depletion'' model \cite{Gierer1972a,Meinhardt2008} (see Fig.~\ref{fig:activator-inhibitor}c). It differs from the activator--inhibitor model in making a specific choice of the reaction terms and reinterpreting the rapidly diffusing component (formerly the inhibitor) as a substrate that is depleted by conversion into the activator.
In this interpretation the autocatalytic \emph{production} (which increases activator and inhibitor concentrations) is replaced by an autocatalytic \emph{conversion} of substrate to activator, which could be understood as membrane attachment of a cytosolic protein. 
However, this type of model \cite{Gierer1972a,Meinhardt2008} crucially depends on cytosolic production of the substrate and degradation of the activator on the membrane. In particular, concentration minima are not the result of a depleted cytosol (as one might expect intuitively), but arise from the dominance of the activator degradation, which effectively suppresses the autocatalytic conversion process \cite{Gierer1972a,Meinhardt2008} (Fig.~\ref{fig:activator-inhibitor}c). In other words, accumulation of cytosolic proteins on the membrane is suppressed by concomitant degradation of their membrane-bound forms. 

Obviously, this assumption is highly specific and biologically implausible in terms of intracellular protein dynamics. The use of a metalanguage with terms like ``activator'' and ``depletion'' suggests that these concepts account for intracellular protein dynamics where finite cytosolic particle pools play an important role. But the draining of finite reservoirs is not actually the mechanism that drives pattern formation in activator--depletion models. Like the activator--inhibitor model the activator--depletion model strictly depends on production and degradation processes to explain pattern formation, and hence, it cannot account for pattern formation in mass conserving systems.
These issues clearly demonstrate that heuristic interpretations and reinterpretations of specific mathematical models do not generalize Turing's insight in a useful way. Indeed, ``activator--inhibitor'' and ``activator--depletion'' models do not even provide a general picture of two-component systems, and two-component systems are already a gross simplification of the biological reality. 
Hence, there is no reason to assume that an intracellular pattern forming system must contain activators and inhibitors, or involve depletion of substrates. In our opinion, the use of such metalanguage to describe the results of quantitative theoretical models does more harm than good. It suggests that a unifying theoretical understanding is provided by the idealized mathematical models to which intuitive terms like ``activator'', ``inhibitor'', and ``depletion'' refer, whereas in reality, little is known about the actual general principles of actual intracellular pattern formation.

Activator--inhibitor models do provide a legitimate phenomenological description of systems based on production (e.g. growth or gene regulation) and degradation, which is applicable to some developmental phenomena \cite{Kondo:2010a} or vegetation patterns \cite{Tarnita:2017a,Rietkerk:2008a}.
However, even in theses cases, it can be argued \cite{Tarnita:2017a} that such models should be integrated in a more complete modelling framework to account for specific details alone, rather than being treated as paradigmatic models that convey the essence of pattern formation in general.

In this article, we provide a molecular perspective on intracellular pattern formation, and review the underlying quantitative biological models, without reference to concepts like ``activators'', ``inhibitors'', or depleting substrates. Instead, we review in the following the specific implementation of pattern forming mechanisms by various protein interaction systems -- i.e. the Min system in \textit{Escherischia coli}, the Cdc42 system in \textit{Saccharomyces cerevisiae}, and the PAR system in \textit{Caenorhabditis elegans}.
Based on these systems we will then extract and discuss recurring principles of the pattern forming dynamics, in particular the fact that intracellular protein dynamics are based on cycling between different conformational states.
We conclude that intracellular pattern formation is, in essence, a spatial redistribution process. Cytosolic concentration gradients are the primary means by which directed transport is facilitated. The establishment and maintenance of such gradients is the key principle underlying self-organized pattern formation. The proper theoretical framework to study intracellular pattern formation is set by mass-conserving reaction-diffusion systems. We will review recent theoretical advanced in this field \cite{Halatek_Frey:2018} at the end of this article.


\subsection{The Min system in \textit{E. coli}}

Cell division in \textit{E. coli} requires a mechanism that reliably directs the assembly of the Z-ring division machinery (FtsZ) to midcell \cite{Lutkenhaus2007a}. How cells solve this task is one of the most striking examples for intracellular pattern formation: the pole-to-pole Min protein oscillation \cite{Raskin1999a}. In the past two decades this system has been studied extensively both experimentally \cite{Hu_etal:1999,Hu_etal:2002,Szeto_etal:2002, Lackner_etal:2003, Mileykovskaya_etal:2003, Hu_Lutkenhaus:1999, Hu_Lutkenhaus:2001, Touhami_etal:2006, Park_etal:2011, Schweizer_etal:2012, Loose2008a, Wu2015, Wu2016} and theoretically \cite{Huang2003a, Fange_Elf:2006, Halatek2012, Hoffmann:2014a, Wu2016, Halatek_Frey:2018}.

The Min protein system consists of three proteins, MinD, MinE, and MinC.
In its ATP bound form the ATPase MinD associates cooperatively with the cytoplasmic membrane (see Fig.~\ref{fig:biochemical-networks}a). Membrane-bound MinD forms a complex with MinC, which inhibits Z-ring assembly. Thus, to form a Z-ring at midcell, MinCD complexes must accumulate in the polar zones of the cell but not at midcell. 
The dissociation of MinD from the membrane is mediated by its ATPase Activating Protein (AAP) MinE, which is also recruited to the membrane by MinD, forming MinDE complexes. MinE triggers the ATPase activity of MinD initiating the detachment of both MinD-ADP and MinE. Subsequently, MinD-ADP undergoes nucleotide exchange in the cytosol such that its ability to bind to the membrane is restored (see Fig.~\ref{fig:biochemical-networks}a).

The joint action of MinD and MinE gives rise to oscillatory dynamics: MinD accumulates at one cell pole, detaches due to the action of MinE, diffuses, and accumulates at the opposite pole. The oscillation period is about 1 min, and during that time almost the entire mass of MinD and MinE is redistributed through the cytosol from one end of the cell to the other and back.

This example nicely illustrates the fact that pattern forming protein dynamics are in essence protein redistribution processes \cite{Halatek2012,Halatek_Frey:2018}. In other words, the emergent phenomenon is directed transport, and not localized production and degradation (depletion), which serve as the basis of activator--inhibitor (or activator--depletion) models.

It was suggested that binding of MinE to the membrane is essential for self-organized pattern formation \cite{Schweizer_etal:2012,Park_etal:2011,Vecchiarelli_etal:2016}. However, theses results were critically debated in the literature \cite{Halatek:2014a} and more recent experiments \cite{Kretschmer:2017a} have explicitly confirmed that MinE membrane binding is not required for self-organized pattern formation. Therefore, we will not discuss this process any further. 

Furthermore, we note that Min protein oscillations are highly regular and therefore amenable to a deterministic description. Instances where stochastic effects \cite{Fischer-Friedrich:2010a} were reported turned out to be overexpression artifacts \cite{Sliusarenko:2011a} and not an indication for intrinsic noise due to low copy numbers.

A striking lesson to be learned from the study of Min protein dynamics is the dependence of the pattern forming process on cell geometry \cite{Raskin1999a,Corbin_etal:2002,Shih_etal:2005,Halatek2012,Varma_etal:2008,Wu2015,Wu2016}. The pole-to-pole oscillation in itself is a phenomenon intrinsically tied to the cell's geometry, which facilitates the detection of a specific location in the cell. Over the past two decades a plethora of fascinating observations has been made:
(i) In filamentous cells, in which cell division is inhibited, the pole-to-pole oscillation develops additional wave nodes showing that the Min oscillation is a standing wave \cite{Raskin1999a}. 
(ii) Experiments with nearly spherical cells show that, in the majority of cases, the pattern forming process is able to detect the long axis, even though it is much less pronounced than in wild-type, rod-shaped cells  \cite{Corbin_etal:2002,Shih_etal:2005}.
(iii) In mutant cells that were grown in nanofabricated chambers of various shapes a broad range of patterns has been observed \cite{Wu2015,Wu2016}. In rectangular cells the oscillation can align with the long axis or the short axis for the same dimensions of the cell. This shows that patterns with distinct symmetries are stable under the same conditions: Min patterns are multistable.


\subsection{The Cdc42 system in \textit{S. cerevisiae}}

Budding yeast (\textit{S. cerevisiae}) cells are spherical and divide asymmetrically by growing a daughter cell from a localized bud. The GTPase Cdc42 spatially coordinates bud formation and growth via its downstream effectors. To that end, Cdc42 must accumulate within a restricted region of the plasma membrane (a single Cdc42 cluster) \cite{Johnson:1999a}. Formation of a Cdc42 cluster, i.e. cell polarization, is achieved in a self-organized fashion from a uniform initial distribution even in the absence of spatial cues (symmetry breaking) \cite{Chant:1991a}.

\begin{figure}
	\includegraphics{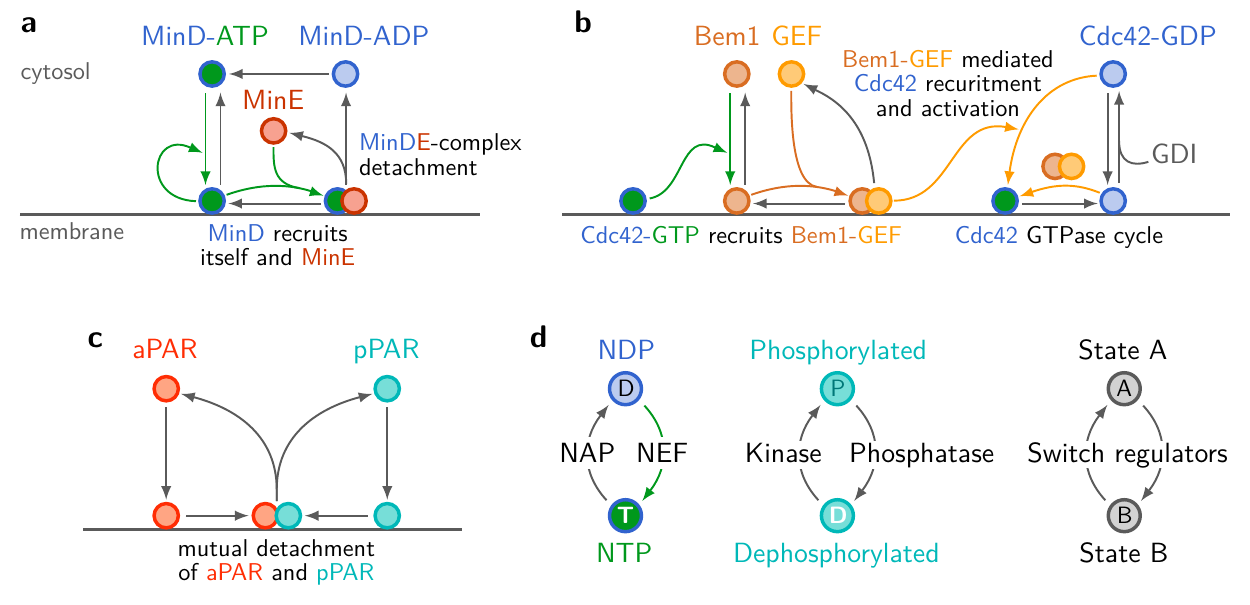} 
	\caption{Biochemical interaction networks of three model systems for self-organized intracellular pattern formation. (a) The Min system of \textit{E. coli} \cite{Huang2003a,Halatek2012}. (b) Cdc42 system of \textit{S. Cerevisiae} \cite{Freisinger:2013a,Klunder:2013a} . (c) PAR system of \textit{C. elegans} \cite{Goehring2011a}. (d) Switching between two conformal states of the proteins involved is a recurring theme in the biochemical networks (a-c). Cycling between membrane bound and cytosolic states is driven by the ATPase/GTPase cycle of MinD and Cdc42 respectively, while the PAP-proteins each cycle between different phosphorylation states. In general we expect switching between distinct conformal states -- catalyzed by ``switch regulators'' such as NTPase-activating proteins (NAPs), nucleotide exchange factors (NEFs), phophatases, and kinases -- to be a core element of biochemical networks that mediate intracellular pattern formation.}
	\label{fig:biochemical-networks}
\end{figure}

Like all other GTPases, Cdc42 switches between an active GTP-bound state, and an inactive GDP-bound state. Both active and inactive Cdc42 forms associate with the plasma membrane, with Cdc42-GTP having the higher membrane affinity. Furthermore, Cdc42-GDP is preferentially extracted from the membrane by its Guanine Nucleotide Dissociation Inhibitor (GDI) Rdi1, which enables it to diffuse in the cytoplasm (see Fig.~\ref{fig:biochemical-networks}b) \cite{Koch:1997a, Johnson:2009a}. Switching between GDP- and GTP-bound states is catalyzed by two classes of proteins: Guanine nucleotide Exchange Factors (GEFs) catalyze the replacement of GDP by GTP, switching Cdc42 to its active state; GTPase Activating Proteins (GAPs) enhance the slow intrinsic GTPase activity of Cdc42, i.e. hydrolysis of GTP to GDP \cite{Bi:2012a} (Note that owing to their biochemical role, GAPs are called activating proteins, even though they switch GTPases into their inactive, GDP-bound state. Moreover, these ``activating proteins'' are in no way related to ``activators'' in the sense of ``activator--inhibitor'' models.). Cdc42 in budding yeast has only one known GEF, Cdc24, and four GAPs: Bem2, Bem3, Rga1, and Rga2. Further, a key player of the Cdc42 interaction network is the scaffold protein Bem1 which is recruited to the membrane by Cdc42-GTP, and itself recruits the GEF (Cdc24) to form a Bem1--GEF complex (Fig.~\ref{fig:biochemical-networks}b) \cite{Butty:2002a, Bose_etal:2001}.
 
Establishment and maintenance of Cdc42 polarization has been shown to rely on two distinct and independent pathways of Cdc42 transport: (i) vesicle trafficking of vesicle-bound Cdc42 along actin cables which require Cdc42-GTP (via its downstream effector Bni1) \cite{Bi:2012a, Slaughter:2009a}, and (ii) diffusive transport of GDI bound Cdc42 in the cytosol. 
Cytosolic Cdc42 is recruited to the membrane by Bem1--GEF complexes (Fig.~\ref{fig:biochemical-networks}b) \cite{Butty:2002a, Bose_etal:2001, Kozubowski:2008a}. Either of these transport pathways is sufficient for viability, as has been shown by either suppressing vesicle trafficking (by depolymerizing actin) or inhibiting cytosolic diffusion of Cdc42-GDP (by knocking out the GDI Rdi1) \cite{Freisinger:2013a, Slaughter:2009a, Marco_etal:2007}. 
Since both pathways depend on Cdc42-GTP, the pattern formation mechanism in both cases relies on polar activation (nucleotide exchange) of Cdc42 by Bem1--GEF complexes, which are in turn recruited by Cdc42-GTP \cite{Witte:2017a, Rapali:2017a, Woods:2015a,Kozubowski:2008a,Irazoqui:2003a}. Various computational models of the Cdc42-Bem1-GEF interaction network confirm that a positive feedback loop mediated by Bem1 is able to establish and maintain polarization \cite{Klunder:2013a, Goryachev:2008a}.

Replacing Cdc42 with a constitutively active mutant suppresses GTPase cycling of Cdc42 and hence restricts it to membranes \cite{Wedlich-Soldner:2003a,Slaughter:2009a}. 
Such mutants show that self-amplified directed vesicle trafficking of Cdc42-GTP provides a viable self-organized polarization mechanism in itself \cite{Wedlich-Soldner:2003a,Wedlich-Soldner:2004a}. Because the mutant Cdc42 is locked in its active state, these cells can forego the polar activation of Cdc42 by Bem1--GEF complexes. Conceptual computational models confirm that an actin mediated transport of Cdc42-GTP can in principle  maintain polarity \cite{Freisinger:2013a, Muller:2016a, Hawkins:2009a, Altschuler:2008a, Wedlich-Soldner:2003a}, although studies of more realistic models show that key details of the involved processes -- endocytosis, exocytosis, and vesicle trafficking -- are still unclear \cite{Layton:2011a, Savage:2012a}. 

Interestingly, experiments where Bem1 was knocked out (or deprived of its ability to recruit the GEF to active Cdc42) in cells with wild-type Cdc42 revealed that a third polarization mechanism must exist, which is independent of both, Bem1 and vesicle trafficking \cite{Smith:2013a,Laan:2015a}. Furthermore, polarization in the complete absence of Cdc42 transport has also been observed \cite{Bendezu:2015a}, hinting at yet another pattern forming mechanism encoded within the interaction network of Cdc42. How these mechanisms operate independently of Bem1-mediated feedback remains an open question that awaits experimental and theoretical analysis.

Normal cell division of budding yeast requires the reliable formation of a single bud-site, i.e. a single Cdc42 cluster (polar zone, sometimes also called ``polar cap''). Various mutant strains exhibit initial transient formation of multiple Cdc42 clusters \cite{Caviston:2002a, Knaus:2007a}, which then compete for the limited total amount of Cdc42, leading to a ``winner-takes-all'' scenario where only one cluster remains eventually \cite{Howell:2009a, Wu:2013a, Witte:2017a}.

\subsection{The PAR system in \textit{C. elegans}}

So far we have discussed examples for intracellular pattern formation in unicellular prokaryotes (Min) as well in eukaryotes (Cdc42). A well studied instance of intracellular pattern formation in multicellular organisms is the establishment of the anterior-posterior axis in the \textit{C. elegans} zygote \cite{Goehring2011a,Hoege2013,Munro_etal:2004,Goehring:2014}. The key players here are two groups of PAR proteins: The aPARs, PAR-3, PAR-6, and aPKC  (atypical protein kinase C) localize in the anterior half of the cell. The pPARs, PAR-1, PAR-2, and LGL, localize in the posterior half. 
In the wild type, polarity is established upon fertilisation by cortical actomyosin flow oriented towards the posterior centrosomes, in other words by active transport of pPAR proteins \cite{Munro_etal:2004,Goehring:2011a}. 
After polarity establishment this flow ceases, but polarity is maintained. In addition, it has been shown that polarity can be established without flow \cite{Goehring:2011a}. These results suggest that PAR protein polarity in \textit{C. elegans} is based on a reaction-diffusion mechanism.

The protein dynamics are based on the antagonism between membrane-bound aPAR and pPAR proteins, mediated by mutual phosphorylation which initiates membrane detachment at the interface between aPAR and pPAR domains near midcell (see Fig.~\ref{fig:biochemical-networks}c).
Thus, PAR-based pattern formation is driven by (mutual) detachment where opposing zones come into contact, and is therefore quite different than the attachment (recruitment) based systems discussed above. 

Despite these apparent differences we will argue in the following sections that patterns formation in all three systems is based on the same general principles. 

\section{General biophysical principles of intracellular pattern formation}

Let us take a bird's eye view and ask: What are the general concepts and recurring themes that are common to pattern formation in all of these biological systems? 

In all cases the biological function associated with the respective pattern is mediated by membrane-bound proteins alone, in other words: \emph{intracellular patterns are membrane-bound patterns} (exceptions are discussed further below). Furthermore, the diffusion coefficients of membrane-bound proteins are generically at least two orders of magnitude lower than those of their cytosolic counterparts, e.g. a typical value for diffusion along a membrane would be between $0.01\; \mu m^2/s$ and $0.1\; \mu m^2/s$, while a typical cytosolic protein has a diffusion coefficient of about $10\; \mu m^2/s$, e.g. see \cite{Bendezu:2015a, Meacci_etal:2006}. 

The key unifying feature of all protein interaction systems is switching between different protein states or conformations. The conformation (state) of a protein can change as a consequence of interactions with other biomolecules (lipids, nucleotides, or other proteins). Likewise, the interactions available to a protein are determined by  its conformation. This can be summarized as the switching paradigm of proteins (Fig. \ref{fig:biochemical-networks}d), which is best exemplified for NTPases such as MinD or Cdc42 whose dynamics are in essence driven by deactivation and reactivation through nucleotide exchange. The phosphorylation and dephosphorylation of PAR proteins by kinases and phosphatases, respectively, exemplifies the same principle. In all these cases, switching is tied to membrane affinity, and thus to the flux of proteins into and out of the cytosol.

Dynamics based on conformational switching conserve the copy number of the protein. Therefore, intracellular protein dynamics are generically represented by mass-conserving reaction-diffusion systems -- and pattern formation in a mass-conserving system can only be based on transport (redistribution), it cannot depend on production or degradation of proteins. In the absence of active transport mechanisms (such as vesicle trafficking) the only available transport process is molecular diffusion. 
Given that membrane-bound proteins barely diffuse, we can assert that the biophysical role of the cytosol in these systems is that of a (three-dimensional) `transport layer'. 
Hence, the (functionally relevant) membrane-bound protein pattern must originate from redistribution via the cytosol, i.e. the coupling of membrane detachment in one spatial region of the cell to membrane attachment in another region, through the maintenance of a diffusive flux in the cytosol.  

However, transport by diffusion eliminates concentration gradients. Hence, if a diffusive flux is to be maintained, a gradient needs to be sustained. Note that due to fast cytosolic diffusion, this  gradient can be rather shallow and still induce the flux necessary to establish the pattern (the flux is simply given by the diffusion coefficient times the gradient).

Intracellular pattern formation is the localized accumulation of proteins on the membrane by cytosolic redistribution. In this context, self-organization is the emergence of directed transport that manifests itself as the formation of spatially separated attachment and detachment zones, due to the interplay between cytosolic diffusion and protein interactions (reactions). 

Furthermore, we note that patterns can also be bound to other structures such as the nucleoid \cite{Lutkenhaus2007a} or the cytoskeleton \cite{Graf:2017a,Yochelis:2015a, Pinkoviezky:2017a}. In several of these pattern forming systems, the dynamics of a nucleoid-bound, ParA-like ATPase results in different patterns, e.g. midcell localization \cite{Schumacher:2017a, Schumacher:2017b} and pole-to-pole oscillations on the nucleoid of a single cargo as well as equidistant positioning of multiple cargoes \cite{Ringgaard:2009a}. Various mechanisms to explain these patterns have been proposed \cite{Howard:2010a}, including models that require ParA filament formation \cite{Ringgaard:2009a, Gerdes_etal:2010, Banigan:2011a} and ones which are based on a self-organizing concentration gradient of the ATPase along the nucleoid \cite{Sugawara:2011a, Surovtsev:2016a, Hu:2017a, Murray:2017a, Walter:2017a}. In all these cases the dynamics of patterns are based on the conformational switching of proteins between the cytosolic and the nucleoid-bound (slowly diffusing) state. For midcell localization, the diffusive flux of the ATPase on the nucleoid was found to be important \cite{Ietswaart:2014a,Schumacher:2017a,Bergeler:2018a}.

Finally, we want to mention pattern forming systems that are based on a preexisting template. The cell geometry itself is such a template, and macromolecules   (phospholipids or specific proteins) can have preferential affinity to accumulate in regions of specific membrane curvature, e.g. the cell poles \cite{Huang:2010a, Laloux:2013a}. However, theoretical analysis of Min protein dynamics indicates that self-organized pattern formation based on lateral instability is robust against heterogeneities in the membrane \cite{Halatek:2012a}.

The elementary (i.e. most simple) intracellular pattern is cell polarization: the asymmetric accumulation of proteins in a cell. It lacks any intrinsic length scale and merely serves to define a specific region of a cell (anterior/posterior domain, budding site). In the following we will use cell polarity patterns as a paradigm to obtain a mechanistic picture of intracellular pattern formation.

\subsection{Cell polarity: The elementary pattern for intracellular pattern formation}

\begin{figure}
	\includegraphics{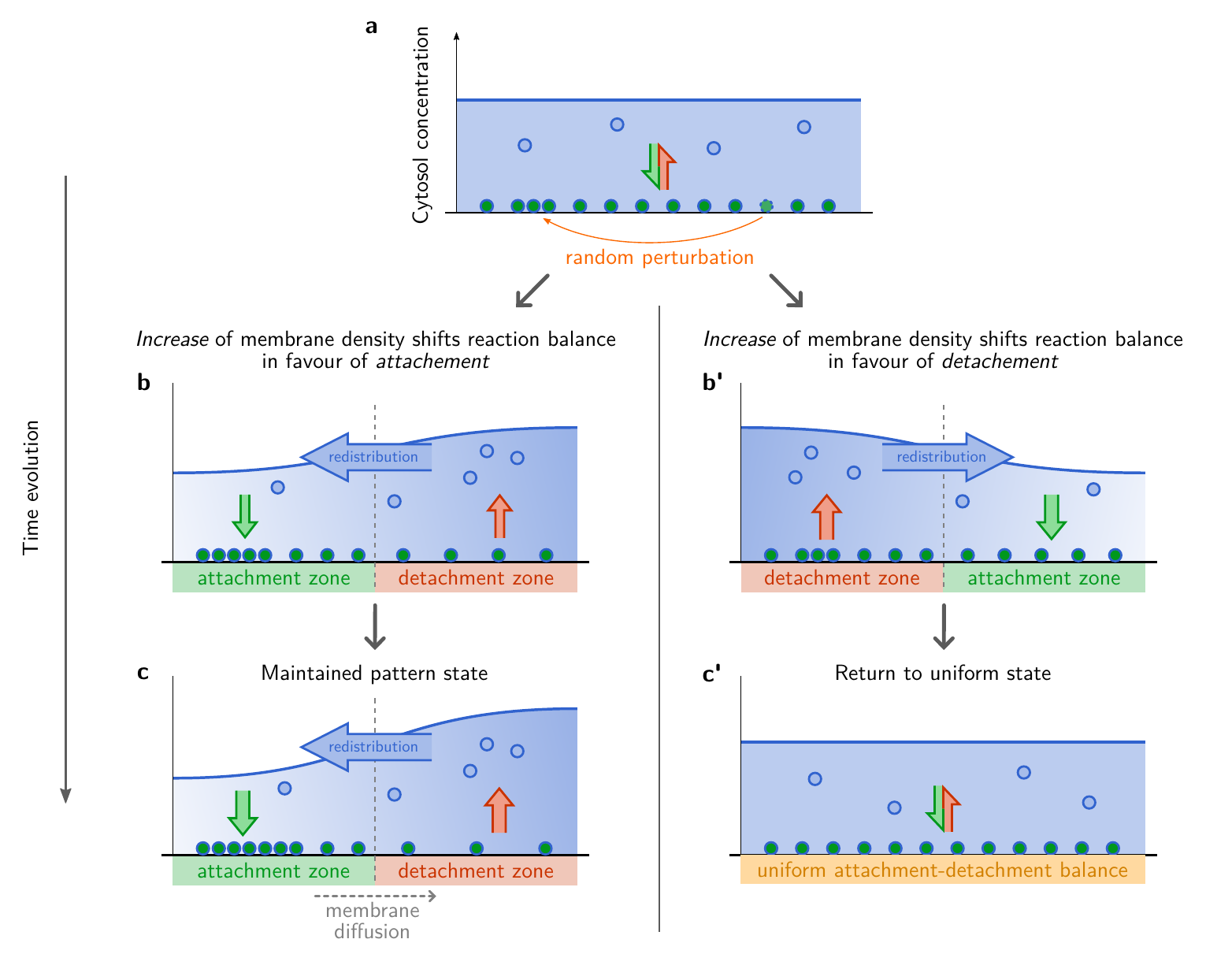}
	\caption{Linear (in)stability of a uniform initial distribution of proteins. (a) In a uniform steady state, attachment and detachment must balance everywhere. An external cue or a random perturbation due to stochastic noise can lead a local increase in membrane density. How the relative balance of attachment and detachment processes shifts in the region of increased membrane density, determines the stability of the uniform state. If the balance in a region of increased membrane density shifts in favour of attachment (b), the region becomes an attachment zone leading to a further increase in membrane density due to redistribution through the cytosol. Hence the spatially separated attachment and detachment zones are maintained, leading the establishment of a pattern. If the balance in a region of increased membrane density shifts the attachment-detachment balance in favour of detachment (b'), this region becomes a detachment zone, while the region of lower membrane density becomes an attachment zone, such that the system returns to its uniform balanced state (c').}
	\label{fig:lateral-stability}
\end{figure}

How can one construct a general conceptual model for self-organized intracellular pattern formation building on the principles discussed in the preceding section?
The protein dynamics are based on cycling of proteins between membrane-bound and cytosolic states, and do not depend on production and degradation of proteins (cf. \cite{Shamir:2016a}). Hence, in a steady state where the protein distribution is spatially uniform, attachment and detachment processes must be balanced throughout the cell.
Lateral instability, e.g. a Turing instability, simply means that small spatial perturbations will be amplified. Let us therefore imagine a perturbation of the density of a membrane-bound species where at some membrane position the protein concentration will be slightly larger than elsewhere on the membrane.
If this perturbation is to be amplified, proteins must be transported to the position where the membrane density is (already) highest. For this transport of proteins, we only have cytosolic diffusion at our disposal. To facilitate a \emph{directed transport} to a specific position by cytosolic diffusion, the cytosolic density at this position must itself be at a minimum  (Fig.~\ref{fig:lateral-stability}). In order to reduce the cytosolic density at the position where the membrane density is highest, the balance between attachment and detachment must shift in favour of attachment such that protein mass flows from the cytosol to the membrane. 
Conversely, in the region where the membrane density is lowest, the attachment-detachment balance must shift in favour of detachment, thereby increasing the cytosolic density in this region.
Only if changes in density shift attachment--detachment balance in this fashion will the initial perturbation on the membrane be amplified by further attachment and detachment due to cytosolic transport (Fig.~\ref{fig:lateral-stability}). 

To concisely summarize: \emph{Intracellular pattern formation can be understood as the formation of attachment and detachment zones, which are coupled through cytosolic gradients that facilitate protein mass redistribution.}
  
In this light, we will now look again at the specific biological systems introduced earlier, and attempt to uncover the basic molecular mechanisms that lead to the formation and maintenance of attachment and detachment zones.
\section{Quantitative models for intracellular pattern forming systems}

\subsection{The Cdc42 system in \textit{S. cerevisiae}}

\begin{figure}
	\includegraphics{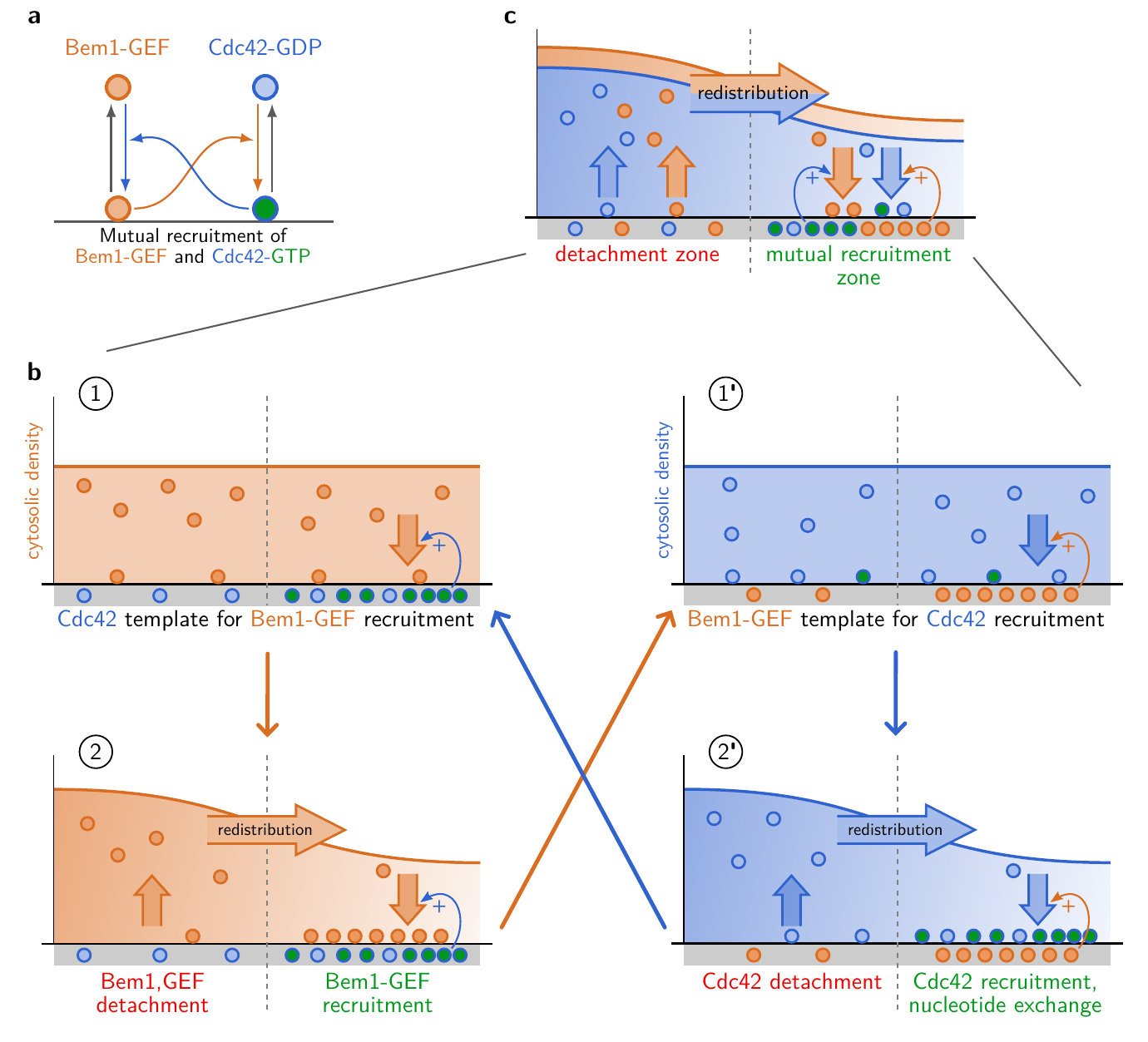} 
	\caption{Mutual recruitment of Cdc42 and Bem1--GEF complexes is the core mechanism of Cdc42 polarization. (a) The interaction network of Cdc42, Bem1 and GEF (Cdc24) can conceptually be simplified to two key processes: Bem1--GEF complexes on the membrane recruit Cdc42-GDP from the cytosol, which is followed by immediate nucleotide exchange (conversion to Cdc42-GTP). Reversely, Cdc42-GTP recruits Bem1 to the membrane by, where it immediately recruits Cdc24 to form Bem1--GEF complexes. (b) A local accumulation of Cdc42-GTP thereby acts as recruitment template for Bem1--GEF complexes (1), creating an attachment zone for Bem1--GEF, while detachment of Bem1--GEF dominates in zones of low Cdc42-GTP density (2). Bem1--GEF complexes then accumulate in their attachment zone. This accumulation acts as a recruitment template for Cdc42 (2), creating co-polarized attachment zones of Cdc42 and Bem1--GEF (1') and (2'). (c) Taken together, the mutual recruitment processes establish and maintain Cdc42-polarization.}
	\label{fig:cdc42-bem1-mechanism}
\end{figure}

The key interaction that drives self-organized Cdc42 polarization is the recruitment of GEF by Cdc42, mediated by Bem1, giving rise to mutual recruitment of Cdc42, Bem1 and GEF. In a minimal model, the role of Bem1 and GEF can be summarized by an effective Bem1--GEF complex, which is recruited to Cdc42-GTP on the membrane (Fig.~\ref{fig:cdc42-bem1-mechanism}a). There, the GEF then recruits more Cdc42 from the cytosol and converts it into the GTP-bound form. As a further simplification, recruitment and nucleotide exchange of Cdc42 by Bem1--GEF complexes can be subsumed into a single step. Similarly hydrolysis of GTP (catalyzed by GAPs) and extraction of CDC42 from the membrane can be conflated to a single membrane dissociation step. Effectively, only active Cdc42 on the membrane and inactive Cdc42 in the cytosol are considered (Fig.~\ref{fig:cdc42-bem1-mechanism}a).

How can this interaction scheme of mutual recruitment establish spatially separated attachment and detachment zones for the polarity marker Cdc42? 
Cdc42-GTP on the membrane acts as a ``recruitment template'' for Bem1--GEF complexes: a zone of high Cdc42-GTP density on the membrane creates an attachment zone for Bem1, which in turn creates an attachment zone for the GEF Cdc24, such that effectively Cdc42-GTP creates an attachment zone for Bem1--GEF complexes (cf. panels (1) and (2) in Fig.~\ref{fig:cdc42-bem1-mechanism}b). A region of high Bem1--GEF density on the membrane, in turn, acts as recruitment template for Cdc42, creating a Cdc42 attachment zone, and locally enhances Cdc42 nucleotide exchange leading to increased local Cdc42-GTP density (cf. panels (1') and (2') in Fig.~\ref{fig:cdc42-bem1-mechanism}b). In the absence of Cdc42-GTP, very little Bem1 attaches to the membrane, such that detachment dominates. Similarly, in the absence of GEF, Cdc42 is dominantly inactive, such that membrane extraction by GDI, i.e. detachment of Cdc42, dominates. Starting from a uniform state, a spatial perturbation of either density will establish a mutual recruitment zone, with Cdc42--GTP sustaining the attachment zone for Bem1--GEF, which in turn maintains the recruitment and activation zone of Cdc42 (Fig.~\ref{fig:cdc42-bem1-mechanism}c).

Conceptually, cell polarization need not require two protein species that mutually recruit each other: conceptual theoretical models for cell-polarity involving only two chemical components (effectively describing a single protein type in two conformational states -- membrane-bound and cytosolic) have also been studied \cite{Otsuji:2007a,Altschuler:2008a,Mori2008}. In these models, patters with multiple density peaks show ``winner-takes-all'' coarsening dynamics due to competition for the conserved total mass of proteins \cite{Otsuji:2007a} (cf. the discussion of wavelength selection and mass-conserving reaction--diffusion systems below). 

\subsection{The Min system in \textit{E. coli}}

\begin{figure}
	\includegraphics{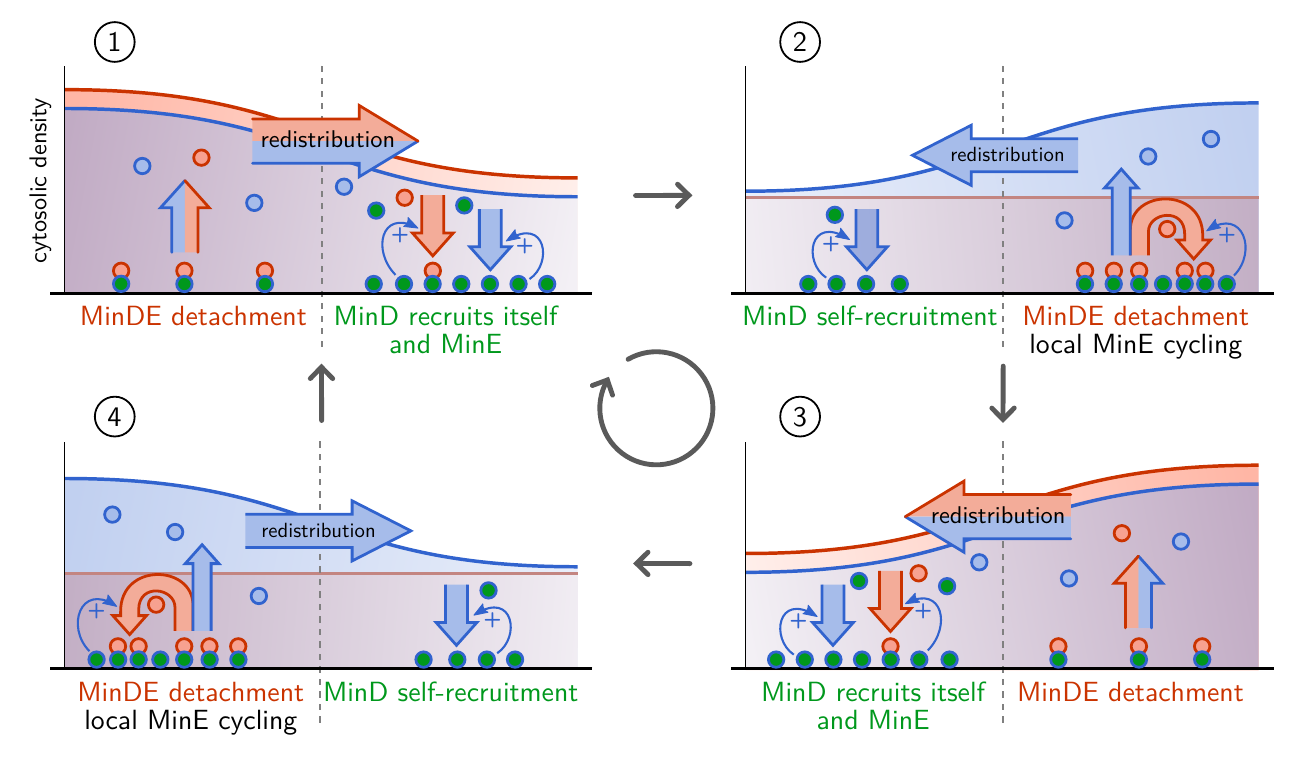} 
	\caption{Pole-to-pole oscillations of MinD and MinE: MinD recruits both itself and MinE from the cytosol creating an attachment zone for both proteins (1). As MinDE complexes accumulate in the polar zone, their detachment begins to dominate over MinD attachment. (2) The old polar zone traps MinE because  it is both an attachment and a detachment zone for MinE, which only cycles locally, as long as there is free MinD left on the membrane. This allows cytosolic MinD to form a new polar zone at the other end of the cell. MinE trapping ends when all MinD has detached from the old polar zone, such that the new polar zone becomes an attachment zone for MinE (3), and the process starts over at the opposite end of the cell (4).}
	\label{fig:min-oscillation-mechanism}
\end{figure}

Pole-to-pole oscillations are the result of interactions between MinD and MinE. In Fig.~\ref{fig:min-oscillation-mechanism}, the key phases of the oscillation cycle are shown. Membrane-bound MinD facilitates further accumulation of MinD and MinE on the membrane through recruitment. The recruitment of MinD shifts the attachment-detachment balance towards further attachment. In contrast, the recruitment of MinE shift this balance towards detachment. 
The structure of a polar zone is such that MinE is accumulated in the form of MinDE complexes at its rim (sometimes called the ``E-ring''), whereas MinD accumulates at its tip (see Fig.~\ref{fig:min-oscillation-mechanism} panels (2) and (4)).

A theoretical analysis of Min protein dynamics revealed that self-organized Min protein pattern formation is based on two requirements \cite{Halatek2012}: The total copy number of MinD must exceed that of MinE, and the recruitment rate of MinE must be larger than that of MinD.
The first condition ensures that all of the MinE can be bound as MinDE complexes on the membrane (MinE-MinD detachment zone, panels (2) and (4) in Fig.~\ref{fig:min-oscillation-mechanism}) while leaving a fraction of MinD free to initiate and maintain a MinD attachment zone.
The second condition causes MinE to be trapped immediately upon entry into a polar zone (MinD-MinE attachment zone) and thereby ``sequestrated'' at the rim, creating a localized MinE-MinD detachment zone (panels (1) and (3) in Fig.~\ref{fig:min-oscillation-mechanism}). Rebinding of MinE to the MinD-MinE attachment zone at the tip of the polar zone is favoured due to the faster recruitment of MinE and the fact that MinD is temporarily inactive after detachment. This leads to the progressive conversion of attachment zones into detachment zones due to the shifting balance in favor of MinDE complexes (detachment). 

Below we will discuss that the inactivation of MinD upon MinE-stimulated hydrolysis is essential for the establishment of intrinsic length scales and for the dependence of the pattern--forming process on cell geometry (see also Fig.~\ref{fig:canalized-transfer}). 

Biochemically the Min system of \textit{E. coli} and the Cdc42 system of \textit{S. cerevisiea}, are closely related: both MinD and Cdc42 are NTPases regulated by enzymatic proteins such as NTPase-activating proteins (NAPs), even though the regulation of Cdc42 activity is much more complex. 
As we have argued above, the pattern forming dynamics of both systems follow the same underlying physical principle: \textit{self-organized spatial separation of attachment and detachment zones}. Indeed, theoretical models have predicted that Min protein dynamics can also give rise to stationary polar patterns \cite{Halatek2012}. 
Conversely, oscillations of Cdc42-marked polar zones in budding yeast have been observed experimentally \cite{Howell:2012a}, while the non-spherical fission yeast exhibits pole-to-pole oscillations of Cdc42 clusters during the polar growth phase \cite{Das:2012a}. This raises the question whether there is a common underlying mechanism that unifies Min protein patterns and Cdc42 polarization at the physical level. 

\subsection{The PAR system in \textit{C. elegans}}

PAR protein polarization in \textit{C. elegans} is based on an antagonism between membrane-bound aPAR and pPAR protein through mutual phosphorylation, cf. \cite{Goehring:2011a}.
The major difference relative to the above discussed systems is the lack of an evident biochemical mechanism for the formation of attachment zones, such as  recruitment. 
Instead for \textit{C. elegans}, attachment zones result from \emph{mutual exclusion}. In a zone with high aPAR concentration on the membrane only aPAR can attach, as pPARs are immediately phosphorylated. Similarly, in zones with high pPAR membrane concentration only pPAR can attach. 
At the interface between aPAR and pPAR zones each protein class drives the other off the membrane. 
Hence, the interface acts as detachment zone for both aPAR and pPAR, whereas the anterior pole (aPAR dominant) acts as aPAR attachment zone, and the posterior pole (pPAR) acts as pPAR attachment zone. 
The key to pattern formation is the detachment zone, i.e. the maintenance of the interface. There, aPAR and pPAR domains are actively separated from each other, and cycling between the interface and the respective attachment zones maintains the bi-polar pattern. Theoretical analysis \cite{Goehring:2011a} shows that a stable interface requires the rates of the antagonistic interactions to be comparable. 

A very interesting aspect of PAR protein pattern formation is the role of the cortical flow \cite{Munro_etal:2004,Goehring:2011a}. In the wild type it is used to segregate aPAR zones from pPAR zones, i.e. to form the respective attachment zones. The polarized state is maintained after the flow ceases, showing that maintenance of the interface is independent of the flow, i.e. it is self-organized. 

\section{Advanced intracellular pattern formation: wavelength selection, dependence on cell geometry, and multistability of patterns}

\subsection{Pattern formation and length scale selection: the classical picture}

Traditionally the phenomenon of self-organized pattern formation in reaction-diffusion systems has been intrinsically linked to the postulated existence of a \emph{characteristic length scale} \cite{Cross1993}. In particular, most authors define a \emph{Turing pattern} as a pattern with a characteristic length scale \cite{Yang:2006a}. In our discussion so far such a length scale has only been mentioned in passing as a phenomenon observed in specific \textit{E. coli} mutants, and in budding yeast mutants as a transient pattern of multiple Cdc42 clusters. Indeed, the theoretical analysis of all quantitative models \cite{Halatek2012,Klunder:2013a,Goehring2011a} discussed so far reveals that the existence of such a length scale is in no way generic -- despite the fact that all patterns emerge from a lateral instability induced by diffusive coupling, i.e. a \emph{Turing instability}.
On the contrary, it appears that the phenomenon biologists refer to as ``the winner takes all'' and physicist as ``coarsening'' is the generic case, cf. \cite{Otsuji_etal:2007,Semplice:2012a, Halatek2012,Wu2016}. Hence, the generic pattern is a polarized state with a single concentration maximum and a single concentration minimum for each chemical component irrespective of the system size.

In fact, the question of length scale selection is a highly nontrivial problem, and can only be addressed in general by numerical simulations.  Linear stability analysis (as introduced by Turing) only predicts the length scale of the (transient) pattern that initially forms from the uniform state (see Fig. \ref{fig:activator-inhibitor}a).  This should not be confused with the length scale of the final pattern.
For instance, coarsening dynamics (``the winner takes all'') are generic examples for dynamics where the length scale collapses to the system size (or an intermediate length scale) regardless of the initially selected length scale \cite{Berry:2018a}.
A case where the length scale is predicted correctly by the linear stability analysis is when the growth of the pattern saturates at small amplitude \cite{Cross1993}. Some authors include saturation at small amplitude in their definition of a Turing pattern \cite{Yang:2002a}. While this definition is mathematically rigorous, it is also very restrictive: For technical mathematical reasons this (supercritical bifurcation, near threshold) case implies that the pattern must vanish if some system parameters are slightly changed. If the pattern does not saturate at a very small amplitude early on, no (reliable) prediction about the final pattern can be made based on the linear stability analysis (except that some non-uniform pattern exists).
 
From the mathematical point of view this specific case (supercritical bifurcation, near threshold) is very attractive as it lends itself to analytical calculations \cite{Cross1993}.
To readers less interested in the mathematical details, these points may seem overly technical. However, it is crucial to realize that -- from the biological perspective -- such technical limitations (a pattern with very small amplitude (weak signal) that is highly sensitive to parameter changes) imply patterns that are highly fragile (cf. \cite{Maini:2012a}), and will therefore be eliminated by natural selection.
In that light it is not surprising that (robust) quantitative models of biological systems, like those presented above, do not meet the constraints of small amplitude and vicinity to a supercritical bifurcation. Partly because most mathematically motivated work on pattern selection is based on the assumption that these constraints are met, very little is known about pattern selection in (evolutionarily robust) biological systems. 

\begin{figure}
	\includegraphics{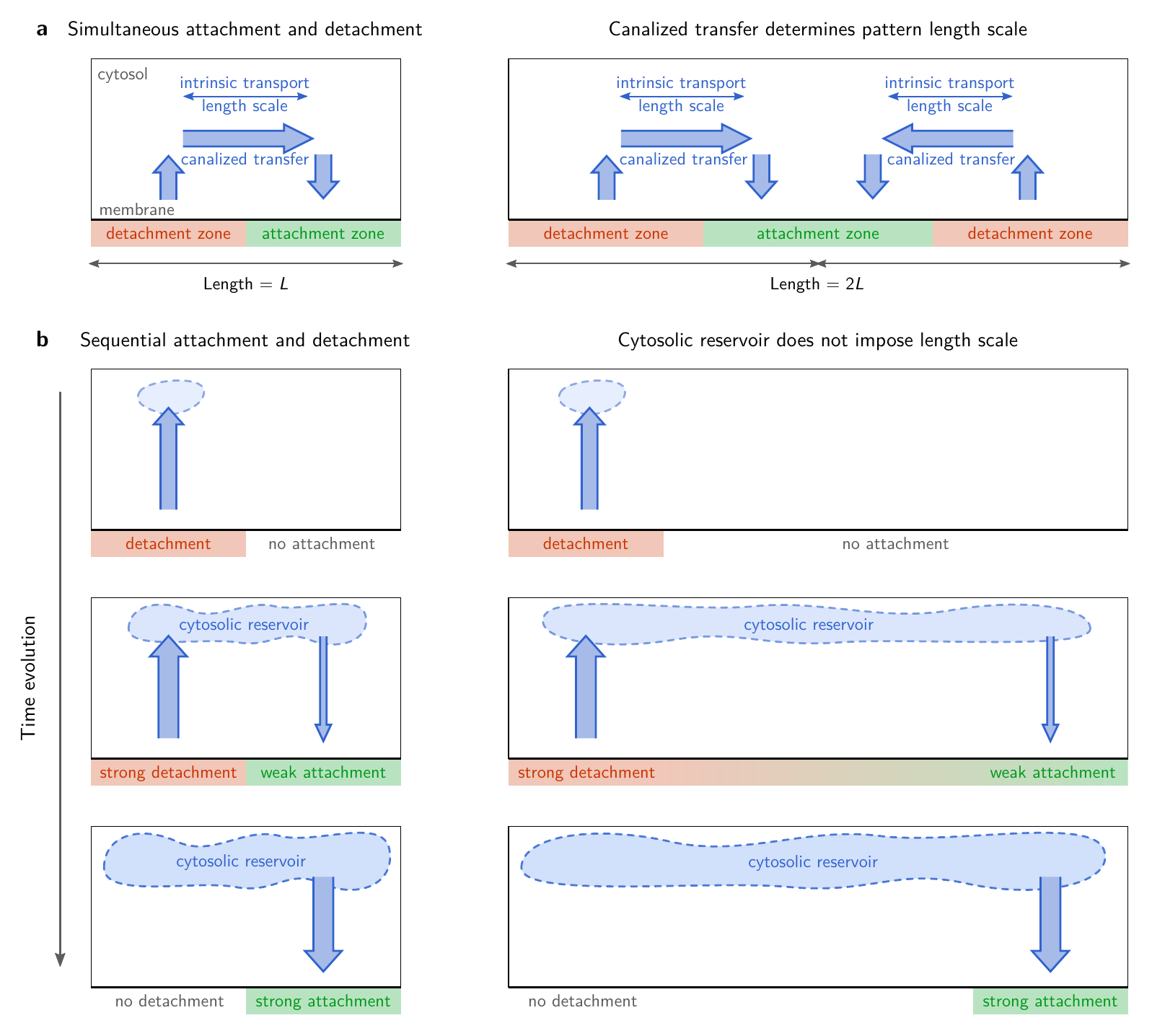} 
	\caption{Canalized transfer of MinD imposes a length scale intrinsic to the reaction--diffusion dynamics in the model developed by Halatek and Frey \cite{Halatek2012}. (a) Canalized transfer refers to the case where attachment-flux and detachment-flux are of similar magnitude such that the cytosolic density does not vary (left). To maintain the flux between detachment and attachment zone a cytosolic gradient must be maintained. The length scale of this gradient dictates the distance between attachment and detachment zones independently of system size. Thus, canalized transfer can give rise to patterns with an intrinsic wavelength in large systems (right). (b) If a detachment zone forms when no attachment zone is available, the detaching proteins fill up a cytosolic reservoir. Once the reservoir reaches a critical density an attachment zone will form. In this case no length scale between attachment and detachment zones is set, since both zones exchange proteins through a (uniform) cytosolic reservoir.}
	\label{fig:canalized-transfer}
\end{figure}

Yet, several key aspects of pattern selection can be inferred from the theoretical analysis of models for actual biological systems. For example, theoretical analysis of Min protein patterns \cite{Halatek2012} showed that standing wave patterns with a finite wavelength emerge if the lateral redistribution of Min proteins is \emph{canalized}, see Fig.~\ref{fig:canalized-transfer}. In terms of the general principles discussed above, this means that attachment and detachment zones are strongly coupled through cytosolic transport. The flux off the membrane in a detachment zone is of the same order of magnitude as the flux onto the membrane in the attachment zone. Hence, the fraction of cytosolic proteins remains approximately constant during the redistribution process (Fig.~\ref{fig:canalized-transfer}a). As we have discussed, attachment and detachment zones are regulated by the membrane kinetics of the specific biochemical model, while transport depends on cytosolic kinetics and diffusion. 
Canalized transfer leads to the emergence of a characteristic separation distance between attachment and detachment zones which depends in a nontrivial manner on the system parameters (Fig.~\ref{fig:canalized-transfer}a). The particular parameter dependence of such characteristic redistribution length scales remains an open question for the Min system, and even more so for general reaction diffusion systems. 
However, it was demonstrated that the total mass flux due to canalized transfer can be inferred from the linear stability analysis for the Min model \cite{Halatek2012}. 
The flux coupling (cytosolic exchange) between detachment and attachment zones is weak, if a cytosolic reservoir is filled and depleted during detachment- and attachment-dominant phases, respectively (see Fig.~\ref{fig:canalized-transfer}b). 
It seems intuitive that a redistribution process through a ``well-mixed'' cytosolic reservoir does not dictate an intrinsic length scale for pattern forming dynamics. 
Moreover, the analysis of quantitative models (such as the Min model) does provide strong evidence that length scale selection in reaction-diffusion systems essentially relates to the length scales of directed (``canalized'') transport.
However, the precise details underlying the emergence of intrinsic length scales remain unknown.

\subsection{Cell geometry and pattern selection}

In the previous section we discussed why it is important to consider the seemingly technical limitations underlying some mathematical results about pattern formation in order to correctly understand quantitative biological models for intracellular pattern formation. 
Besides the question of length scale selection, the effect of system geometry (i.e. cell shape) is in this respect another case in point.

At first, linear stability analysis of reaction-diffusion systems was exclusively restricted to planar geometries such as lines and flat surfaces.
Only very recently, the method was extended to account for (2d) circular geometries where dynamics can take place on the boundary of the circle (membrane) as well as in the bulk (cytosol), and proteins exchange between the two domains (membrane-cytosol cycling) \cite{Levine_Rappel:2005}. This important first step towards quantitative modeling of intracellular protein dynamics was still limited to purely linear attachment-detachment dynamics (thus excluding the cases of cooperative attachment, recruitment, or antagonistic detachment). 
Later, linear stability analysis methods were extended to general attachment-detachment kinetics by for (3d) spherical geometry \cite{Klunder:2013a}, and for (2d) elliptical geometry \cite{Halatek2012}.

The extension to elliptical geometry revealed a very important point \cite{Halatek2012}: it is in no way generic that patterns align with the long axis of a cell, i.e. there is in general no intrinsic preference for the selection of long axis patterns over short axis patterns.
From a biological perspective this is a crucial finding, since proper axis selection is typically linked to the spatial nature of the biological function mediated by the pattern in the first place.
Intuitively, one may expect that axis selection is connected to the ``characteristic length scale'' of the pattern obtained from a linear stability analysis in a planar (flat) geometry: whichever axis length of the cell is closer to this ``characteristic length scale'' determines the axis that is selected. The intuition behind this is that the pattern has ``to fit'' into the cell. So far, however, there is no evidence to support this intuition. On the contrary, it appears that the question of axis selection is much more complicated.
In a study combining theory and experiments Wu et al. \cite{Wu2015,Wu2016} analyzed the Min protein patterns in rectangular cell geometries of various sizes and aspect ratios. 
The experiments found that a broad range of patterns (aligned with the long axis or the short axis) can emerge in the same system geometries. Hence, intracellular Min protein patterns are multistable, and can conform to a variety of intrinsic length scales instead of one ``characteristic'' length scale.

Theoretical analysis \cite{Wu2016} confirmed multistability of Min patterns, and was able to link all observations to the emergence of an intrinsic length scale for diffusive cytosolic redistribution (``canalized transport'') in the model: the stronger the flux-coupling between attachment and detachment zones, the stronger was the dependence of the pattern-forming process on cell geometry, and the greater the number of multistable patterns with distinct symmetries (long axis or short axis alignment) observed in a broad range of rectangular cell geometries \cite{Wu2016}. This strongly suggest that pattern selection and the influence of cell geometry are -- just like wavelength selection -- closely tied to the length scale of lateral transport and the strength of the coupling between attachment and detachment zones.
Note that this is a very different picture from the one presented by ``activator--inhibitor'' models. In the latter, the length scale is set by the degradation and production length scale, e.g. the length scale over which autocatalytic production of the slowly diffusing component (activator) dominates over its own degradation. In contrast, for intracellular protein dynamics the essential length scale appears to be the (mean) distance over which membrane-bound proteins (the slowly diffusing components) are redistributed in the (fast diffusing) cytosolic state following their detachment. 

We will next discuss how the switch-like behaviour of proteins appears to be essential for the emergence and regulation of such transport length scales.

\subsection{The different roles of cytosolic kinetics and membrane kinetics}

As we discussed above, the switch-like behaviour of proteins between active and inactive states is a central paradigm of protein dynamics.
In many cases the switch alters the affinity of proteins for the membrane and can thus be utilized to stimulate attachment or detachment. In case of the Min system in \textit{E. coli} only active MinD-ATP can bind and be recruited to the membrane. As theoretical analysis \cite{Halatek2012} has shown, this property is essential for the regulation of intracellular transport and the establishment of ``canalized transfer'': Since MinD detaches from the membrane in the inactive MinD-ADP form it cannot immediately rebind. Hence, if the timescale of cytosolic reactivation (nucleotide exchange) is sufficiently long, a MinD protein detaching from a polar zone with high MinD membrane density can leave the polar zone, by diffusion, before being reactivated. 
In this way, rebinding of detached MinD to polar zones can be suppressed even if the affinity of cytosolic MinD-ATP for membrane-bound MinD is very high. 
In fact, a very high affinity for membrane-bound MinD can serve to promote rebinding of MinD-ATP in new (weak) polar zones. In other words: to promote directed transport of MinD from high (MinD and MinE membrane) density region to a low density region. In this way, the high density region becomes a detachment zone and the low density region an attachment zone. Increasing the affinity of cytosolic MinD-ATP to membrane-bound MinD (recruitment rate) simply increases the attachment in the low density region -- but not in the high density region where MinD detached and cannot rebind due to delayed nucleotide exchange. Hence, the coupling (mass flux) between detachment and attachment zones increases with the MinD recruitment rate, which leads to ``canalized transfer''.

A recent study \cite{Thalmeier2016} has also shown that the interplay between a membrane affinity switch and cell geometry can lead to an entirely new type of intracellular patterning that is not based on lateral instabilities (such as the Turing instability) or excitability. In this case the uniform steady state does not become laterally unstable but is \emph{replaced} by a non-uniform pattern, i.e. it ceases to exist. The experimental observation \cite{Zhang:2009a} is that the MinD homolog AtMinD from \textit{Arabidopsis thaliana} forms a bipolar pattern in $\Delta\text{MinDE}$ \textit{E. coli} cells.
There is evidence that the ATPase AtMinD can bind (cooperatively) to the membrane in both, its ADP and its ATP form. Assuming that AtMinD detaches from the membrane in the ADP bound form, the theoretical analysis shows that the membrane distribution of AtMinD in steady state is always non-uniform (bi-polar) if, (i) the membrane affinities of AtMinD-ADP and AtMinD-ADP are different, and (ii) the geometry of the cell deviates from a spherical  shape. 
The mechanism underlying such \emph{geometry induced pattern formation} is based on the \emph{local ratio of membrane surface area to cytosolic volume}: in an elliptical cell geometry, a protein detaching from a cell pole  is more likely to re-encounter the membrane in unit time than a protein that detaches from a site closer to midcell. By utilizing different membrane affinities (of ADP and ATP states) and cytosolic switching (between these states) a protein system can then establish highly robust bipolar patterns that reflect the cell's geometry. Again, they key process underlying pattern formation is cytosolic redistribution -- in this case combined with geometry-dependent reattachment.

\section{Mass-conserving reaction diffusion models: A new paradigm?}
In the preceding sections we emphasized that mass conservation is the major unifying property of intracellular pattern-forming protein dynamics. Over the past decade, mass-conserving reaction-diffusion systems received considerable attention in the theoretical literature \cite{Otsuji:2007a, Ishihara:2007a, Pogan:2012a, Goh:2011a, Mori2008, Mori2011, Semplice:2012a, Kessler:2016a, Halatek_Frey:2018}. These studies try to answer how mass-conservation affect reaction-diffusion dynamics. But what is the relevance  for biological systems? 

In \cite{Mori2008} a mass-conserving model for cell polarity is proposed -- where mass-conservation leads to the halting of a propagating wave front. This ``pinned'' wave represents the polarized pattern. 
It has been argued that the corresponding pattern-forming mechanism is not related to a Turing instability but instead based on excitability and bistability \cite{Mori2008,Mori2011}. A similar line or reasoning is presented in \cite{Semplice:2012a}. However, it has been pointed out recently \cite{Goryachev:2017b, Trong2014a} that a Turing instability in the ``wave-pinning'' model is recovered upon parameter change. 
Other studies report that mass-conserving reaction diffusion systems are prone to coarsening \cite{Otsuji:2007a, Ishihara:2007a}. However, it remains elusive whether a general relation between mass-conservation and coarsening exists.

Recently, it has been shown that the general mechanism of pattern formation in mass-conserving reaction diffusion systems is based on the lateral redistribution of the conserved quantities \cite{Halatek_Frey:2018}. The total amount of conserved quantities (protein copy number) determines the position and stabilities of chemical equilibria. Spatiotemporal redistribution of conserved quantities shifts local chemical equilibria and is generically induced by any lateral instability with unequal diffusion coefficients (such as the Turing instability). The pattern forming dynamics simply follow the movement of local equilibria and the final patterns are scaffolded by the spatial distribution of local equilibria. This study further demonstrated that ``wave-pinning'' patterns originate from the same physical processes as Turing instabilities: the redistribution of conserved quantities and the shifting local chemical equilibria.  
In future research it will be interesting to see how the formation of attachment and detachment zones can be formalized within the mass-redistribution framework.

\section{Summary and Discussion}

It should be clear by now that activator--inhibitor models do not provide the appropriate  concepts to account for intracellular pattern formation.
Rather, the generic key feature of pattern forming protein system is conformational switching. Proteins cycle between different states such as active and inactive, or membrane-bound and cytosolic. It is the switching of (conformational) states that drives the system and leads to pattern formation, not the production and degradation of proteins, which is the basis of any activator--inhibitor or activator--depletion model.  
Any dynamics based on the switching (conformational) states conserved the total copy number. Therefore, the proper models for intracellular protein dynamics are mass-conserving reaction-diffusion systems.

In any mass-conserving system pattern formation has to be based on redistribution of mass. The question to be asked about self-organization in such systems is how redistribution comes about, i.e. how directed transport is established and maintained.

In the context of intracellular protein pattern formation there is a clear functional division between membrane-bound and cytosolic protein distributions: The biologically (functionally) relevant pattern is that of the membrane-bound factor(s), while the cytosol acts as a transport medium which facilitates the formation and maintenance of the membrane-bound pattern.
The basis of intracellular pattern formation is therefore the emergence of spatially separated \emph{attachment and detachment zones} and their coupling (transport from detachment to attachment zone) through cytosolic gradients.
The key design principle for pattern forming protein networks is therefore the ability to set up and maintain attachment and detachment zones in the presence of ongoing protein redistribution. 

We have discussed how pattern formation in three different biological systems is facilitated by the formation of attachment and detachment zones. In budding yeast, cell polarity is established by localized accumulation of Cdc42 on the inner face of the plasma membrane. Pattern formation is based on the localized formation of mutual attachment zones for Cdc42-GTP and Bem1--GEF complexes through mutual recruitment. In \textit{E. coli}, pole-to-pole oscillations emerge due to the interactions of MinD and MinE. This process is based on the formation of attachment zones with high MinD membrane density due to the recruitment of MinD and MinE from the cytosol. If the balance in a polar zone is shifted towards higher MinE/MinD ratios, an attachment zone becomes a detachment zone. The sequestration of MinE in detachment zones enables the formation of new attachment zones some distance away. The conversion of attachment zones to detachment zones and vice versa by the slow  shift in the MinE/MinD ratio within a zone is the basis for the oscillation. The establishment of the anterior-posterior axis in \textit{C. elegans} is based on PAR protein polarization. Here, pattern formation originates from the formation of a (mutual) detachment zone near midcell, and attachment zones exclusive for pPAR or aPAR, respectively, at the two cell poles. The establishment and maintenance of this pattern requires that the rates of both antagonistic processes are balanced. 

The question about length scale selection and pattern selection in general is still open. Apart from mathematically idealized cases that do not apply to biological systems no general statements about the wavelengths of patterns can be made. 
However, theoretical studies of Min protein pattern formation suggest that the emergence and selection of finite wavelengths is closely tied to the simultaneous formation and diffusive coupling of attachment and detachment zones. A key step in the regulation of the diffusive coupling between attachment and detachment zones is the cytosolic switching between conformation with high and low affinity for the membrane (cytosolic nucleotide exchange). Strikingly, there is evidence that such cytosolic switching processes play a key role in mediating the sensitivity of self-organized pattern formation to cell geometry \cite{Wu2016}. 

Many key questions about intracellular pattern formation remain open. In our opinion, the focus on concepts based on activator--inhibitor models in the discussion of pattern formation phenomena has been a hindrance to progress rather than a help.  

Intracellular pattern forming protein dynamics are most generally expressed by mass-conserving reaction-diffusion systems. Local equilibria, as recently suggested \cite{Halatek_Frey:2018}, are a promising candidate for a general and unifying theoretical framework for such systems. To advance our understanding of intracellular protein dynamics a theoretically rigorous analysis of pattern forming instabilities in mass-conserving reaction-diffusion systems would be highly desirable. In his seminal article Turing presented the general idea of pattern forming instabilities in reaction-diffusion systems. Since its publication 65 years ago only little has been learned about the general physical principles underlying the \emph{Turing instability}. We expect that a focus on the mass-conserving case will finally enable us to extract some general physical principles of pattern formation systems based on Turing's lateral instability.  

\section{Acknowledgments}
This research was funded by the German Excellence Initiative via the NanoSystems Initiative Munich, and by the Deutsche Forschungsgemeinschaft (DFG) via Project B02 within SFB 1032 (Nanoagents for Spatio‐Temporal Control of Molecular and Cellular Reactions), Areas A and C within GRK2062 (Molecular Principles of Synthetic Biology), and Project P03 within TRR174 (Spatiotemporal Dynamics of Bacterial Cells). The authors  thank S.~Bergeler and W.~Daalman for feedback on the manuscript.

\clearpage

\clearpage

\clearpage

\clearpage

\clearpage

\clearpage

\begin{figure}
	\includegraphics{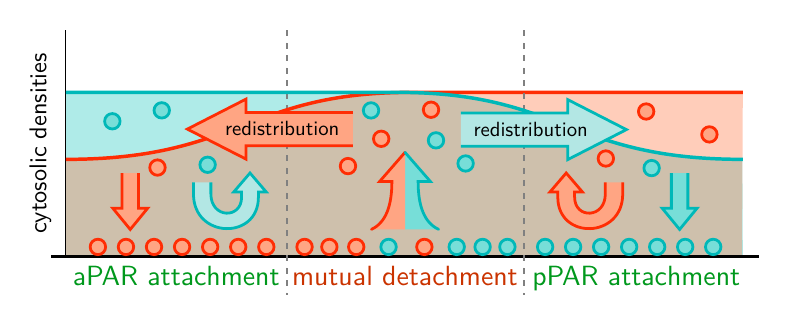} 
	\caption{Mutual antagonism between aPAR and pPar proteins creates a detachment zone at the interface between aPAR- and pPAR-dominated regions. A region of high aPAR density on the membrane is a detachment zone for pPAR, such that detaching pPARs can only attach to a region of low aPAR density, and vice versa. A balance of the mutual antagonistic processes is necessary to prevent one protein species from dominating over the other and taking over the whole membrane.}
	\label{fig:par-mechanism}
\end{figure}

\clearpage

\end{document}